\documentclass{article}
\usepackage{graphicx,amssymb,amsmath}
\renewcommand{\d}{\mathrm{d}}
\def\beq{\begin{eqnarray}}
\def\eeq{\end{eqnarray}}
\def\beqs{\begin{equation}\begin{split}}
\def\eeqs{\end{split}\end{equation}}
\begin{document}

\title{Bianchi models with vorticity: The type III bifurcation}
 \author{A A Coley and S Hervik\thanks{Address after April 15, 2008: Dept. of Mathematics and Natural Sciences, University of Stavanger, N-4036 Stavanger, Norway.} \\ \\
Department of Mathematics \& Statistics,\\
 Dalhousie University,\\
Halifax, Nova Scotia, \\
Canada B3H 3J5 \\
(\tt {aac, herviks@mathstat.dal.ca})  }

%\thanks{}%
%\subjclass{}%
%\keywords{}%

%\date{\today}%

\maketitle
\begin{abstract}
We study the late-time behaviour of tilted perfect fluid Bianchi type III models using a dynamical systems approach. We consider models with dust, and perfect fluids stiffer than dust, and eludicate the late-time behaviour by studying the centre manifold which dominates the behaviour of the model at late times. In the dust case, this centre manifold is 3-dimensional and can be considered as a double bifurcation as the 2 parameters ($h$ and $\gamma$) of the type VI$_h$  model are varied. We calculate the decay rates and show that for dust or stiffer the models approach a vacuum spacetime, however, it does so rather slowly: $\rho/H^2\sim 1/\ln t$. 
\end{abstract}
%\begin{keywords} 
%Center manifolds, asymptotic expansions, dynamical systems in physics.
%\end{keywords}

%\begin{AMS}
%34E05, 34K09, 37N20
%\end{AMS}

\pagestyle{myheadings}
\thispagestyle{plain}
\markboth{A. COLEY AND S. HERVIK}{THE TYPE III BIFURCATION}

\section{Introduction} 

In previous papers the dynamical behaviour of tilted Bianchi models have been studied \cite{Prev,CH2,HLim,HHLC2,BHtilted} using the dynamical systems approach \cite{DynSys}. In particular, the tilted Bianchi models of type VI$_h$ was studied in \cite{HHLC2}, where it was pointed out that the type III=VI$_{-1}$ model
is a bifurcation value in terms of the group parameter $h$. The analysis of this model therefore requires a more complicated centre manifold analysis. 
In addition, in terms of the equation of state parameter $\gamma$ there is also a bifurcation at $\gamma=1$, which results in a 3-dimensional centre manifold in this case\footnote{The 3rd dimension of this centre manifold is related to the existence of a line of equilibrium points and is present even for non-tilted models.}. The dust type III model is therefore a particularly interesting model as it appears as a double bifurcation (this was also pointed out in \cite{BHtilted}). Due to the centre manifolds arising from these bifurcations, the Bianchi type III models with $1\leq \gamma< 2$ was not studied in detail in \cite{HHLC2}. 

The type III model also have an interesting geometric interpretation. The type III Lie group can be considered as the Thurston geometry $\mathbb{H}^2\times \mathbb{R}$ which plays an important role in 3-dimensional geometry \cite{thurston}. Interestingly, this Thurston geometry also permits for a type VIII action which allows for a possible connection between the behaviours of these models. In \cite{HHLC2} such a connection was pointed out;  however,  the oscillatory behaviour present in the type VIII models \cite{HLim} have no analogue in the type III models.

The tilted Bianchi type III model is the last of the ever-expanding Bianchi models  left to study in terms of its late-time behaviour. It is the aim of this paper to complete this analysis. Some partial results are known; for example, the case $0<\gamma<1$ was analysed in \cite{HHLC2} (including an analysis in some particular subspaces). Here, we will consider the fully tilted type III model and show that for $1\leq \gamma <2$ the late-time asympotote is the self-similar vacuum spacetime given by
\[ \d s^2=-\d t^2+t^2(\d x^2+e^{-2x}\d y^2)+\d z^2.\]
The tilt, on the other hand, depends on $\gamma$ in the following way: for $\gamma=1$, the tilt tends to zero (albeit slowly), while for $1<\gamma<2$ the tilt is asymptotically extreme. 

The dynamical system considered in this paper is a prime example of a dynamical system appearing in other theories of applied mathematics and physics. This dynamical system, which is also constained, is of relatively high dimension, namely 7.  The analysis performed here is a certain aspect of this dynamical system, namely a centre manifold analysis. This centre manifold appears as two of the parameters of the theory experience bifurcation values. We apply centre manifold theory, and for all the variables of the theory, we find the decay rates and the correction terms, as the dynamical time $\tau\rightarrow \infty$.  

\section{Equations of motion}
%\subsection{The orthonormal frame approach}
%The line-element of a Bianchi cosmology can be written
%\beq
%\d s^2=-\d t^2+\delta_{ab}{\mbold\omega}^a{\mbold\omega}^b,
%\eeq 
%where $t$ is the co-moving cosmological time. The one-forms ${\mbold\omega}^a$ are left-invariant one-forms on the hypersurfaces spanned by the group orbits. 
%The geometric (or normal) congruence, $n^{\mu}$, is given by ${\bf n}=\partial/\partial t$. It is also useful to define the shear and the Hubble scalar associated with the congruence $n^\mu$: 
%\beq
%H\equiv\frac 13n^\mu_{~;\mu}, \quad \sigma_{\mu\nu}\equiv n_{\mu;\nu}-H h_{\mu\nu}, 
%\eeq
%where $h_{\mu\nu}$ is the spatial metric on the hypersurfaces spanned by the group orbits. 
%The matter variables are chosen to be the energy density, $\mu$, and the tilt-velocity, $v^a$, which is defined as the 3-velocity of the fluid with respect to the geometric (or normal) congruence, $n^{\mu}$. 
%The equations of motion can now be written down in terms of the Hubble scalar, $H;$ the shear, $\sigma_{ab}$; the curvature variables $n^{ab}$ and $a_c$; and the matter variables $\mu$ and $v^a$.  

%In the dynamical systems approach it is common to introduce expansion-normalised variables (we divide the variables with the appropriate powers of $H$).   
The dimensionless variables of the theory are the shear variables, $(\Sigma_+,\Sigma_-,\Sigma_{12},\Sigma_{13},\Sigma_{23})$; geometric variables, $(A,N,\lambda)$; and fluid variables $(\Omega,v_1,v_2,v_3)$. The variable $\Omega$ is the (expansion-normalised) energy-density while $v_i$ are so-called tilt variables (they correspond to the velocity of the fluid). If $0<V<1$, where $V^2=v_1^2+v_2^2+v_3^2$, we call the model tilted; the special cases $V=0$ and $V=1$ are  called non-tilted and extremely tilted, respectively. 
Moreover, the dimensionless time-variable, $\tau$, can be related to proper time, $t$, by 
\[ \frac{\d \tau}{\d t}=H, \]
where $H$ is the Hubble scalar. 

The papers \cite{CH2,HHLC2} contain all the details regarding the  determination of the evolution 
equations for the models  under consideration.

The equations of motion for the general tilted type III model
are:
% (see \cite{CH2,HHLC2} for the complete derivation of the equations): 
\beq 
\Sigma_+'&=& (q-2)\Sigma_++{3}(\Sigma_{12}^2+\Sigma^2_{13})-2N^2 +\frac{\gamma\Omega}{2G_+}\left(-2v_1^2+v_2^2+v_3^2\right) \\
\Sigma_-'&=&(q-2-2\sqrt{3}\Sigma_{23}\lambda)\Sigma_-+\sqrt{3}(\Sigma_{12}^2-\Sigma_{13}^2) +2AN +\frac{\sqrt{3}\gamma\Omega}{2G_+}\left(v_2^2-v_3^2\right)\\
\Sigma'_{12}&=& \left(q-2-3\Sigma_+-\sqrt{3}\Sigma_-\right)\Sigma_{12} -\sqrt{3}\left(\Sigma_{23}+\Sigma_-\lambda\right)\Sigma_{13} +\frac{\sqrt{3}\gamma\Omega}{G_+}v_1v_2\\
\Sigma'_{13}&=&\left(q-2-3\Sigma_++\sqrt{3}\Sigma_-\right)\Sigma_{13}-\sqrt{3}\left(\Sigma_{23}-\Sigma_-\lambda\right)\Sigma_{12}+\frac{\sqrt{3}\gamma\Omega}{G_+}v_1v_3\\
\Sigma'_{23}&=&(q-2)\Sigma_{23}-2\sqrt{3}N^2\lambda+2\sqrt{3}\lambda\Sigma_-^2+2\sqrt{3}\Sigma_{12}\Sigma_{13}+ \frac{\sqrt{3}\gamma\Omega}{G_+}v_2v_3\\
N'&=& \left(q+2\Sigma_++2\sqrt{3}\Sigma_{23}\lambda\right){N}\\
\lambda' &=& 2\sqrt{3}\Sigma_{23}\left(1-\lambda^2\right)\\
A'&=& (q+2\Sigma_+)A . 
\eeq 
The equations for the fluid are:
\beq
\Omega'&=& \frac{\Omega}{G_+}\Big\{2q-(3\gamma-2)+2\gamma Av_1
 +\left[2q(\gamma-1)-(2-\gamma)-\gamma\mathcal{S}\right]V^2\Big\}
 \quad \\
 v_1' &=& \left(T+2\Sigma_+\right)v_1-2\sqrt{3}\Sigma_{13}v_3-2\sqrt{3}\Sigma_{12}v_2-A\left(v_2^2+v_3^2\right)-\sqrt{3}N\left(v_2^2-v_3^2\right)\\
 v_2'&=& \left(T-\Sigma_+-\sqrt{3}\Sigma_-\right)v_2-\sqrt{3}\left(\Sigma_{23}+\Sigma_-\lambda\right)v_3+\sqrt{3}\lambda{N}v_1v_3+\left(A+\sqrt{3}N\right)v_1v_2 \\
 v_3'&=& \left(T-\Sigma_++\sqrt{3}\Sigma_-\right)v_3-\sqrt{3}\left(\Sigma_{23}-\Sigma_-\lambda\right)v_2-\sqrt{3}\lambda{N}v_1v_2+\left(A-\sqrt{3}N\right)v_1v_3 \\
 V'&=&\frac{V(1-V^2)}{1-(\gamma-1)V^2}\left[(3\gamma-4)-2(\gamma-1)Av_1-\mathcal{S}\right],
\eeq 
where 
\beq q&=& 2\Sigma^2+\frac
12\frac{(3\gamma-2)+(2-\gamma)V^2}{1+(\gamma-1)V^2}\Omega\nonumber \\
\Sigma^2 &=& \Sigma_+^2+\Sigma_-^2+\Sigma_{12}^2+ \Sigma_{13}^2+\Sigma_{23}^2\nonumber \\
\mathcal{S} &=& \Sigma_{ab}c^ac^b, \quad c^ac_{a}=1, \quad v^a=Vc^a,\quad \nonumber \\
 V^2 &=& v_1^2+v_2^2+v_3^2,\quad  \nonumber \\
 T&=& \frac{\left[(3\gamma-4)-2(\gamma-1)Av_1\right](1-V^2)+(2-\gamma)V^2\mathcal{S}}{1-(\gamma-1)V^2}\nonumber\\
 G_+&=&1+(\gamma-1)V^2\nonumber.
\eeq 
These variables are subject to the constraints 
\beq
1&=& \Sigma^2+A^2+N^2+\Omega \label{const:H}\\
0 &=& 2\Sigma_+A+2\Sigma_-N+\frac{\gamma\Omega v_1}{G_+} \label{const:v1}\\
0 &=&
-\left[\Sigma_{12}(N+\sqrt{3}A)+\Sigma_{13}\lambda{N}\right]+\frac{\gamma\Omega v_2}{G_+} \label{const:v2}\\
0 &=&
\left[\Sigma_{13}(N-\sqrt{3}A)+\Sigma_{12}\lambda{N}\right]+\frac{\gamma\Omega v_3}{G_+} \label{const:v3} \\
0&=& A^2-3\left(1-\lambda^2\right)N^2.\label{const:group} 
\eeq 
In this paper we will be concerned with the case when the parameter $\gamma$ obeys $1\leq \gamma <2$. The case $0<\gamma<1$ is considered in \cite{HHLC2}. 
The generalized Friedmann equation (\ref{const:H}) yields an expression which effectively defines the energy density $\Omega$. We will assume that this energy density is non-negative: $\Omega\geq 0$. 
Therefore, the state vector can thus be considered 
${\sf X}=[\Sigma_+,\Sigma_-,\Sigma_{12},\Sigma_{13},\Sigma_{23},N,\lambda,A,v_1,v_2,v_3]$ 
modulo the constraint equations  (\ref{const:v1})-(\ref{const:group}). 
The evolution takes place on a seven dimensional submanifold of the twelve dimensional space; thus the dimension of the physical state space is seven. 

%Let us investigate the constraint equations in a little more detail. The five constrains constitute a set of equations 
%\[ {\mathcal C}_i({\sf X})=0, \] 
%and in order for these to define a submanifold of the higher-dimensional space, we need to check that the matrix 
%\[ {\sf M}\equiv \left(\frac{\partial\mathcal{C}_i}{\partial X^j}\right)=\begin{bmatrix}
%\frac{\partial\mathcal{C}_1}{\partial \Sigma_+} & \frac{\partial\mathcal{C}_1}{\partial \Sigma_-} & \cdots \\ 
%\frac{\partial\mathcal{C}_2}{\partial \Sigma_+} & \frac{\partial\mathcal{C}_2}{\partial \Sigma_-} & \cdots \\ 
%\vdots & \vdots & \end{bmatrix},\] 
%evaluated on the constraints, has maximal rank. Now it can be shown, that the rank of ${\sf M}$ is maximal in the interior of the tilted type III state space; however, on special sets being part of the boundary it may be lower. An example of this is the very special Bianchi type I case for which $A=N=0$. Fortunately, the set of points where the rank of ${\sf M}$ is reduced, are all invariant subspaces, and hence, such points can be treated separately. Therefore, if the initial value is a point in state space for which the rank of ${\sf M}$ is maximal, the rank of $M$ will at all times remain maximal. 

Additional details are presented in \cite{CH2,HHLC2}.

\subsection{Fluid Vorticity} In general, the fluid will have a non-zero vorticity, $W^{\alpha}$. The vorticity of the fluid for the type III model is given by:
\beq
W_a=\frac{1}{2B}\left(N_{ab}v^b+\varepsilon_{abc}v^bA^c+\frac 1{1-V^2}N_{bc}v^bv^cv_a\right), \quad W_0=-v^aW_a, 
\label{vorticity}\eeq
where 
\[ B\equiv\frac{1-\frac 13(V^2+V^2\mathcal{S}+2A_av^a)}{\sqrt{1-V^2}\left[1-(\gamma-1)V^2\right]}.\]

\subsection{Equilibrium points} 
The equilibrium points of the Bianchi type III model with $1\leq \gamma<2$ are all given in \cite{HHLC2}. We will be concerned with the general case for which the following equilibrium points are of importance. 
\subsubsection{$B(I)$: Equilibrium points of Bianchi type I}  
\begin{enumerate} 
\item{}$\mathcal{I}(I)$: $\Sigma_+=\Sigma_-=\Sigma_{12}=\Sigma_{13}=\Sigma_{23}=A=N=V=0$ and $\Omega=1$. Here, $|\lambda|<1$ and is an unphysical parameter.  
\end{enumerate}

\subsubsection{$T(III)$: Vacuum case ($\Omega=0$)} \label{pp}
All of these equilibrium points are plane wave solutions and have
\beq &&\Omega=\Sigma_{12}=\Sigma_{13}=\Sigma_{23}=0, ~\Sigma_-=N=\sqrt{-\Sigma_+(1+\Sigma_+)},\nonumber 
\\ &&  ~A=(1+\Sigma_+),~ -1<\Sigma_+<0,~ |\lambda|<1.\nonumber 
\eeq 
It is also advantageous to introduce $r\equiv\sqrt{1-\lambda^2}$. This implies that we can write 
\[ \Sigma_+=-\frac{1}{1+3r^2}, \quad 0<r\leq 1. \]  
We will also define $\rho$ by 
\[ \rho=v_2^2+v_3^2.\]

The equilibrium points are then determined by the tilt velocities:
\begin{enumerate}
\item{} $\mathcal{L}(-1)$: $v_1=v_2=v_3=0$. 

\item{} $\widetilde{\mathcal{L}}(-1)$: $v_1=\frac{\gamma(1+3r^2)-2(1+2r^2)}{2r^2(\gamma-1)}$, $v_2=v_3=0$, $\frac{2(1+3r^2)}{1+5r^2}<\gamma<2$. 
\item{} $\widetilde{\mathcal{L}}_{\pm}(-1)$: $v_1=\pm 1$, $v_2=v_3=0$. 
\item{}\label{defF} $\widetilde{\mathcal{F}}^+(-1)$: $ v_1=-\frac{\gamma(1+3r^2)-(1+5r^2)}{2r^2(2-\gamma)}$,  $(v_2^2-v_3^2)= \rho r$,\\  $\rho=\frac{(1+r^2)\left(3-2\gamma\right)\left[\gamma(1+3r^2)-(1+5r^2)\right]}{4r^4(2-\gamma)}$. 
where $
\frac{1+5r^2}{1+3r^2}\leq \gamma \leq \frac{3}{2}$ for $1<r(r+\sqrt{r^2+1})$ and $\frac{1+5r^2}{1+3r^2}\leq \gamma \leq \frac{1+3r^2}{1+r^2}$ for $ r(r+\sqrt{r^2+1})\leq 1$. 
\item{} $\widetilde{\mathcal{F}}^-(-1)$: This case collapses to the line-bifurcation $\gamma=1$, $v_1=0$, $(v_2^2-v_3^2)=-\rho r$, $0<\rho<1$.  

\item{} $\widetilde{\mathcal{E}}_p^+(-1)$: Here, $1>r(r+\sqrt{r^2+1})$ and $ v_1=-\frac{2r^2}{(1-r^2)}$, $(v_2^2-v_3^2)= \rho r$,  
$\rho= 1-v_1^2=\frac{(1+r^2)(1-3r^2)}{(1-r^2)^2}$. 

\item{} $\widetilde{\mathcal{E}}_p^-(-1)$: Here, $-1<r(r-\sqrt{r^2+1})$ and $v_1=0$,  $(v_2^2-v_3^2)=- r$, $\rho=1$.

\end{enumerate}

\subsection{Equilibrium points of special importance} 
The following equilibrium points all have  non-negative eigenvalues and are therefore potential late-time attractors: 
\begin{enumerate}
\item{} $\mathcal{P}(III)$: This is the special case $\left.\widetilde{\mathcal{F}}^-(-1)\right|_{r=1}$ and is a special line-bifurcation of 'tilted' vacuum solutions for $h=-1$, $\gamma=1$:
\beq
&& \Sigma_+=-\frac 14, \quad \Sigma_-=N=\frac{\sqrt{3}}{4}, \quad A=\frac 34, \nonumber \\
&&\Sigma_{12}=\Sigma_{13}=\Sigma_{23}=\lambda=0, \quad v_1=v_2=0,\quad 0<v_3<1. \nonumber
\eeq
The extreme limit of this equilibrium point, $\lim_{v_3\rightarrow 1}\mathcal{P}(III)=\left.\widetilde{\mathcal{E}}_p^-(-1)\right|_{r=1}$. We will call $ \left.\widetilde{\mathcal{E}}_p^-(-1)\right|_{r=1}$ for $\mathcal{E}^-_p(III)$ for simplicity.  We will also define  $\mathcal{P}_0(III)\equiv \lim_{v_3\rightarrow 0}\mathcal{P}(III)=\left.{\mathcal{L}}(-1)\right|_{r=1}$. 

Regarding the eigenvalues; there are 3 eigenvalues which are zero and the rest all have a negative real part. In order to resolve the stability properties of these equilibrium points one has to resort to centre manifold theory.
\item{} $\mathcal{P}_0(III)\equiv\left.{\mathcal{L}}(-1)\right|_{r=1}$, $\gamma=1$.
\item{} $\mathcal{E}^-_p(III)\equiv \left.\widetilde{\mathcal{E}}_p^-(-1)\right|_{r=1}$, $1\leq \gamma<2$.
\end{enumerate}

\section{The late-time behaviour}
In \cite{HHLC2} the eigenvalues of all the equilibrium points were computed. This analysis showed that \emph{the only potential equilibrium points acting as local attractors are $\mathcal{P}(III)$ and $\mathcal{E}^-_p(III)$.} Furthermore, an extensive numerical analysis indicates there are no other kind of attractors (like attracting curves). Lastly, the existence of monotone functions \cite{HHLC2} also restricts the attractors to lie in certain subsets. Although, we have no proof, this indicates that the following local attractors are indeed global attractors (except for a set of measure zero). 

For the $\gamma=1$  line of equilibria $\mathcal{P}(III)$, two of the zero eigenvalues of the linearised system actually correspond to a non-trivial Jordan block; i.e., one of the Jordan blocks is of the form 
\[
{\sf J}_1=\begin{bmatrix} 0 & 1 \\ 0 & 0 \end{bmatrix}.\nonumber
\]
This means that a generic solution `drifts' along the line of equilibria; the amount of `drifting' depends on the second order terms. However, it can be shown that the line $\mathcal{P}(III)$ is indeed unstable; the solutions drift towards the end point $\mathcal{P}_0(III)$. 

\subsection{The case $\gamma=1$: Attractor is $\mathcal{P}_0(III)$.} 
The centre manifold in this case is 3-dimensional. The potential attracting equilibrium point is  $\mathcal{P}_0(III)$. The centre manifold can be found as follows. It is useful to define $\widehat{\Sigma}_+$ and $\widehat{\Sigma}_-$ by rotating $\Sigma_+$ and $\Sigma_-$ according to:
\beq
(\Sigma_+,\Sigma_-)=\left(-\frac 12\widehat{\Sigma}_++\frac{\sqrt{3}}{2}\widehat{\Sigma}_-,\frac{\sqrt{3}}2\widehat{\Sigma}_++\frac1{2}\widehat{\Sigma}_-\right).
\eeq

Let us define the variables $x^i$: 
\[ (\widehat{\Sigma}_+,\Sigma_{23},N,\lambda,v_1,v_2,v_3)=\left(\frac 12+x_1,x_2,\frac{\sqrt{3}}4+x_3,x_4,x_5,x_6,x_7\right).\] 
To determine the late-time behaviour it is actually necessary to expand the variables to 3rd order in ${\sf x}=(x^i)$ (2nd order is not sufficent to determine all the coefficients of the leading order terms of the variables). 
The variables, $\widehat{\Sigma}_-$, $\Sigma_{12}$ and $\Sigma_{13}$ can be found by solving the linear constraints, eqs. (\ref{const:v1}-\ref{const:v3}), while $\Omega$ can be found by solving the Friedmann constraint, eq. (\ref{const:H}). 

The equations can now be written: 
\beq
{\sf x}'={\sf L}{\sf x}+{\sf C}({\sf x},{\sf x})+{\sf T}({\sf x},{\sf x},{\sf x})+\mathcal{O}({\sf x}^4),
\eeq
where ${\sf C}$ and ${\sf T}$ is a bilinear and trilinear vector-valued function, respectively. 

Now, it is advantageous to define ${\sf y}={\sf P}{\sf x}$ so that the linear term in ${\sf y}$ is the Jordan canonical form of ${\sf L}$. This can be accomplished by defining
\beq
{\sf P}=\begin{bmatrix} 
 1 & 0 & 2\sqrt{3} & 0 & 0 & 0 & 0 \\
 1 & 0 &-2\sqrt{3} & 0 & 0 & 0 & 0 \\
-1 & 1 &-2\sqrt{3} & \frac{\sqrt{3}}4 & 0 & 0 & 0 \\
-1 & 1 & 2\sqrt{3} & 0 & 0 & 0 & 0 \\
0  &-1 & 0 & -\frac{\sqrt{3}}{4} & 0 & 0 & 1 \\
0 & -1 & 0 & 0 & 1 & 0 & 0 \\
0 & 0 & 0 & 0 & -1 & 1 & 0
\end{bmatrix}.  
\eeq
In terms of ${\sf y}$ we get 
\beq
{\sf y}'={\sf J}{\sf y}+\widetilde{\sf C}({\sf y},{\sf y})+\widetilde{\sf T}({\sf y},{\sf y},{\sf y})+\mathcal{O}({\sf y}^4),
\eeq
where 
\[ {\sf J}=\mathrm{diag}\left(0,~-\frac 32,~0,-\frac 32, ~0,-\frac 32~,-\frac 32 \right).\] 
From this we note that the 3-dimensional centre manifold can be parameterised by the variables $(x,y,z)\equiv (y_1,y_1+y_3,y_3+y_5)$. 

The next step is to find functions $Y_i(x,y,z)$, $i=2,4,6,7$ so that $y_i-Y_i(x,y,z)=0$ are invariant submanifolds. Since the centre manifold will dominate the late-time behaviour, the variables $y_i$, $i=2,4,6,7$ will behave as $y_i\approx Y_i(x,y,z)$ at late times.

To third order in $(x,y,z)$ we obtain: 
\beq
Y_2(x,y,z)&\approx& -y^2+x(x+z)^2, \nonumber \\
Y_4(x,y,z)&\approx& y^2+\frac 43yx-x(x+z)^2, \nonumber \\
Y_6(x,y,z)&\approx& \frac 43 x[-y+2(x+z)^2], \nonumber \\
Y_7(x,y,z)&\approx&-\frac 23(x+z)[4x(x+z)+\sqrt{3}y].\nonumber
\eeq
Substituting this into the equations for $(x,y,z)$ we finally get the equations on the centre manifold: 
\beq
x'&=&x^2+\frac{1}{3}x^3+10xy^2+\mathcal{O}({\sf y}^4), \nonumber \\
y'&=&2yx-\frac{2}{3}yx^2-6y^3+\mathcal{O}({\sf y}^4), \nonumber \\
z'&=& xz-\frac 13x^3-13xy^2-3y^2z+\mathcal{O}({\sf y}^4).
\eeq
These equations can now be solved to give:
\beq 
x(\tau)&=&-\frac{1}{\tau-\tau_0}-\frac{\ln(\tau-\tau_0)}{3(\tau-\tau_0)^2}+\mathcal{O}[(\tau-\tau_0)^{-2}], \nonumber \\
y(\tau)&=&\frac{C}{(\tau-\tau_0)^2}+\frac{2C\ln(\tau-\tau_0)}{3(\tau-\tau_0)^3}+\mathcal{O}[(\tau-\tau_0)^{-3}], \nonumber \\
z(\tau)&=&\frac{1+v_0}{\tau-\tau_0}+\frac{(1+v_0)\ln(\tau-\tau_0)}{3(\tau-\tau_0)^2}+\mathcal{O}[(\tau-\tau_0)^{-2}], \nonumber 
\eeq
where $\tau_0$, $v_0$ and $C$ are constants.
Substituting the solutions into the original variables, we get the decay rates: 
\beq
\widehat{\Sigma}_+&=&\frac 12\left(1-\frac 1\tau-\frac{1}{3}\frac{\ln \tau}{\tau^2}\right)+\mathcal{O}\left(\tau^{-2}\right), \nonumber \\
\widehat{\Sigma}_-&=&\frac{\sqrt{3}(8v_0^2-3C^2)}{9\tau^4}\left(1+\frac 43\frac{\ln \tau}{\tau}\right) +\mathcal{O}(\tau^{-5}), \nonumber \\
\Sigma_{12}&=& -\frac {4v_0C}{3\tau^4}\left(1+\frac{4}{3}\frac{\ln \tau}{\tau}\right)+\mathcal{O}(\tau^{-5}), \nonumber \\
 \Sigma_{13}&=& \frac {2\sqrt{3}v_0}{3\tau^2}\left(1+\frac{2}{3}\frac{\ln \tau}{\tau}\right)+\mathcal{O}(\tau^{-3}), \nonumber \\
\Sigma_{23}&=& -\frac {4C}{3\tau^3}\left(1+\frac{\ln \tau}{\tau}\right)+\mathcal{O}(\tau^{-4}), \nonumber \\
N&=& \frac{\sqrt{3}}{4}\left(1-\frac{1}{3\tau}-\frac{\ln \tau}{9\tau^2}\right)+\mathcal{O}(\tau^{-3}), \nonumber \\
\lambda&=& \frac{4\sqrt{3}C}{3\tau^2}\left(1+\frac{2}{3}\frac{\ln \tau}{\tau}\right)+\mathcal{O}(\tau^{-3})\nonumber \\
\Omega &=& \frac{1}{\tau}\left(1+\frac 13\frac{\ln\tau}{\tau}\right)+\mathcal{O}(\tau^{-2}), \nonumber \\
v_1&=& -\frac 83\frac{v_0^2}{\tau^3}\left(1+\frac{\ln\tau}{\tau}\right)+\mathcal{O}(\tau^{-4}), \nonumber \\
v_2&=& -\frac{2\sqrt{3}}3\frac{v_0C}{\tau^3}\left(1+\frac{\ln\tau}{\tau}\right)+\mathcal{O}(\tau^{-4}), \nonumber \\
v_3&=& \frac{v_0}{\tau}\left(1+\frac 13\frac{\ln\tau}{\tau}\right)+\mathcal{O}(\tau^{-2}).
\label{decay1}\eeq
 The constant, $\tau_0$, has been eliminated by a shift of time, $\tau$. We have also included the second term in the expansion which is decaying as $(\ln\tau)\tau^{-1}$ compared to the 1st order term, and the order of the correction terms for all the variables. 

The Hubble scalar and the cosmological time can also be calculated to give
\beq
H\approx H_0\tau^{\frac 12}e^{-\frac 32\tau}, \quad t-t_0\approx \frac{2}{3H_0}\tau^{-\frac 12}e^{\frac 32\tau}. 
\eeq

This shows that the attractor is indeed the equilibrium point  $\mathcal{P}_0(III)$. It is therefore plausible that  all solutions, except a set of measure zero, approach this equilibrium point as a power law at late times; in particular, the expansion-normalised energy density goes as $\Omega\propto 1/\tau$ for sufficently large $\tau$. 

Let us consider some other physical consequences of this result. As we now how the decay rates, we can calculate the behaviour of certain physical quantities. Consider, for example, the fluid vorticity given in eq.(\ref{vorticity}). The fluid in this case is pressureless matter (dust). At late times, the (square of) the spatial vorticity of a pressureless matter in a Bianchi type III model, will decay as:
\[ W^aW_a\propto \frac{1}{\tau^2}.\]
Hence, the vorticity decays fairly slowly. Therefore, due to the presence of the centre manifold, certain physical quantities will decay slower than the usual exponential decay seen in most other Bianchi models\footnote{There might be other physical consequences as well, for example in the cosmic microwave background. Such a slow decay of the vorticity will most likely have observational consequences for the cosmic microwave background radiation \cite{Jaffe1}.}.

\subsection{The case $1<\gamma<2$: Attractor is $\mathcal{E}^-_p(III)$.}
For the case $1<\gamma<2$ the potential attractor is the extremely tilted equilibrium point $\mathcal{E}^-_p(III)$. In this case the centre manifold is 2-dimensional so the analysis simplifies slightly. We can simplify even further by restricting to the extremely tilted invariant subspace $V=1$ in which the centre manifold lies. 

The analysis is analogous to the $\gamma=1$ case; however, we can also solve for $v_3=\sqrt{1-v_2^2-v_3^2}$. We again define the variables $x^i$ as follows: 
\[ (\widehat{\Sigma}_+,\Sigma_{23},N,\lambda,v_1,v_2)=\left(\frac 12+x_1,x_2,\frac{\sqrt{3}}4+x_3,x_4,x_5,x_6\right).\] 
We will now expand the equations of motion to 2nd order in ${\sf x}=(x^i)$. The variables, $\widehat{\Sigma}_-$, $\Sigma_{12}$ and $\Sigma_{13}$ can be found by solving the linear constraints, eqs. (\ref{const:v1}-\ref{const:v3}), while $\Omega$ can be found by solving the Friedmann constraint, eq. (\ref{const:H}), as before. 

We again align our variables with the Jordan canonical form, so we define: 
\beq
{\sf P}=\begin{bmatrix} 
 1 & 0 & 2\sqrt{3} & 0 & 0 & 0 \\
 1 & 0 & 2\sqrt{3} & 0 & -\frac 38 & 0 \\
-1 & \frac{4\sqrt{3}}{3} &-2\sqrt{3} & 1 & 0 & 0 \\
-1 & 0 & 1-2\sqrt{3} & 0 & \frac 83 & 0 \\
0  &-\frac{8\sqrt{3}}{9} & \frac 23 & -\frac 23 & 0 & -\frac 43 \\
0  &1+\frac{8\sqrt{3}}{9} & -\frac 53 & \frac 23 & 0 & \frac 43 
\end{bmatrix}
\eeq
The Jordan matrix is 
\[ {\sf J}=\mathrm{diag}\left(0,~-\frac 32,~0,-\frac 32,-\frac 32~,-\frac 32 \right);\] 
hence, we can use $y_1$ and $y_3$ to parameterise our centre manifold. 

A similar procedure as in the dust case gives us the following equation of motion on the centre manifold: 
\beq
y_1'&=&8y_1^2+\mathcal{O}({\sf y}^3), \nonumber \\
y_3'&=&8y_1y_3+\mathcal{O}({\sf y}^3).
\eeq
To lowest order, this can be integrated to yield $y_1=-1/(8\tau)$, $y_3=(v_0+1)/(8\tau)$, where a constant has been eliminated by a shift of time. 
Substituting these expressions into the variables gives the decay rates (the errors are also indicated) : 
\beq
\widehat{\Sigma}_+&=&\frac 12\left(1-\frac 1{4\tau}\right)+\mathcal{O}[(\ln\tau)/\tau^2], \nonumber \\
\widehat{\Sigma}_-&=&-\frac{\sqrt{3}(9v_0^2-128)}{9216\tau^2} +\mathcal{O}[(\ln\tau)/\tau^3], \nonumber \\
\Sigma_{12}&=& -\frac {\sqrt{3}v_0}{192\tau^2}+\mathcal{O}[(\ln\tau)/\tau^3], \nonumber \\
 \Sigma_{13}&=& \frac {\sqrt{3}}{12\tau}+\mathcal{O}[(\ln\tau)/\tau^2], \nonumber \\
\Sigma_{23}&=& -\frac {\sqrt{3}v_0}{48\tau^2}+\mathcal{O}[(\ln\tau)/\tau^2], \nonumber \\
N&=& \frac{\sqrt{3}}{4}\left(1+\frac{(3v_0^2+40)}{768\tau^2}\right)+\mathcal{O}[(\ln\tau)/\tau^3], \nonumber \\
\lambda&=& \frac{v_0}{8\tau}+\mathcal{O}[(\ln\tau)/\tau^2]\nonumber \\
\Omega &=& \frac{1}{8\tau}+\mathcal{O}[(\ln\tau)/\tau^2], \nonumber \\
v_1&=& -\frac 1{3\tau}+\mathcal{O}[(\ln\tau)/\tau^2], \nonumber \\
v_2&=& -\frac{v_0}{16\tau}+\mathcal{O}[(\ln\tau)/\tau^2], \nonumber \\
v_3&=& 1-\frac{9v_0^2+256}{4608\tau^2}+\mathcal{O}[(\ln\tau)/\tau^3] \nonumber \\
\sqrt{1-V^2} &=& \left(\sqrt{1-V^2}\right)_0\tau^{-\frac{(2\gamma-3)}{4(2-\gamma)}}e^{-\frac{3(\gamma-1)}{(2-\gamma)}\tau}.
\label{decayg1}\eeq
We note that also in this case the energy-density approaches zero, but it does so rather slowly: $\Omega\propto 1/\tau$. The tilt, on the other hand, evolves exponentially towards extreme tilt. 

The Hubble scalar and the cosmological time are found to be
\beq
H\approx H_0e^{-\frac 32\tau}\tau^{\frac 18}, \quad t-t_0\approx \frac{2}{3H_0}e^{\frac 32\tau}\tau^{-\frac 18}.
\eeq

From this result we can see that the cosmological time, which corresponds to the proper time of observers following the geometrically defined congruence, is unbounded as $\tau\rightarrow \infty$. If we instead consider the congruence defined by the four-velocity of the fluid \cite{tilt1}, the proper time is given by 
\[ T-T_0=\int^\tau_{\tau_0} \frac{1}{H}\sqrt{1-V^2}\d\tau\sim e^{-\frac{3(3\gamma-4)}{2(2-\gamma)}\tau}.\] 
Interestingly, for $\gamma>4/3$, the observers following the fluid congruence will reach infinite expansion in finite proper time. This is an indication that the fluid experiences some kind of singularity in the future \cite{tilt1}.

\section{Conclusion}
As an application of centre manifold theory, we have studied the centre manifold of the tilted Bianchi type III model. For the cases $\gamma=1$ and $1<\gamma<2$ this centre manifold is 3- and 2-dimensional, respectively. We analysed the late-time evolution and determined the decay rates for the variables (eq.(\ref{decay1}) for $\gamma=1$, and eq.(\ref{decayg1}) for $1<\gamma<2$). This analysis is a  prime example of dynamical systems theory (in particular, centre manifold theory) appearing in applied mathematics and physics. 

\section*{Acknowledgements} 
We would like to thank R.J. van den Hoogen and W.C. Lim for discussions.

\end{document}